\documentclass[apjl]{emulateapj}
\usepackage{latexsym}
\usepackage{epsfig}
\usepackage{natbib}

\shorttitle{Large amplitude longitudinal oscillations}
\shortauthors{M. Luna et al.}

\begin{document}

\title{Large amplitude longitudinal oscillations in a solar filament}

\author{M. Luna\altaffilmark{1}, and J. Karpen\altaffilmark{2}}

\altaffiltext{1}{CRESST and Space Weather Laboratory NASA/GSFC, Greenbelt, MD 20771, USA}
\altaffiltext{2}{NASA/GSFC, Greenbelt, MD 20771, USA}

\begin{abstract}
  We have developed the first self-consistent model for the observed large-amplitude oscillations along filament axes that explains the restoring force and damping mechanism. We have investigated the oscillations of multiple threads formed in long, dipped flux tubes through the thermal nonequilibrium process, and found that the oscillation properties predicted by our simulations agree with the observed behavior. We then constructed a model for the large-amplitude longitudinal oscillations that demonstrates that the restoring force is the projected gravity in the tube where the threads oscillate. Although the period is independent of the tube length and the constantly growing mass, the motions are strongly damped by the steady accretion of mass onto the threads by thermal nonequilibrium. The observations and our model suggest that a nearby impulsive event drives the existing prominence threads along their supporting tubes, away from the heating deposition site, without destroying them. The subsequent oscillations occur because the displaced threads reside in magnetic concavities with large radii of curvature. Our model yields a powerful seismological method for constraining the coronal magnetic field and radius of curvature of dips. Furthermore, these results indicate that the magnetic structure is most consistent with the sheared-arcade model for filament channels.
\end{abstract}

\section{Introduction}\label{intro-sec}

High-cadence H$\alpha$ observations of large-amplitude standing oscillations along a filament axis were first reported by \citet{jing2003}, who measured bulk displacements of up to 140 Mm, with a period of $80\min$ and a high velocity amplitude of $92~\mathrm{km~s^{-1}}$. Prominence oscillations are classified as ``large amplitude'' when velocity is of the order of $20~\mathrm{km~s^{-1}}$ or larger \citep{oliver2002}. Similar large-amplitude longitudinal (LAL) oscillations were observed later by \citet{jing2006} and \citet{vrsnak2007}. All the LAL oscillations reported so far show periods of $0.83-2.66~\mathrm{hours}$, damping times of $2-10~\mathrm{hours}$, damping time per period of $2.3-6.2$, and amplitudes of $30-100~\mathrm{km~s^{-1}}$. These oscillations apparently were triggered by an energetic event: a sub-flare, a microflare, or a flare close to the filament. In addition, prominences undergo large-amplitude transverse oscillations in response to waves emanating from distant flares \citep[e.g.,][]{gilbert2008,tripathi2009}. The work presented here addresses only the longitudinal motions.

Several models have been proposed to explain the LAL oscillations. \citet{jing2006} interpreted their observed oscillations in terms of the \citet{kleczek1969} model. In this model the prominence is considered a slab with a magnetic field along the filament, the restoring force is the magnetic tension, and the motion is perpendicular to the main axis of the filament. However, this geometry is inconsistent with the observed direction of the LAL motion and the most likely magnetic structure: dipped or twisted field nearly parallel to the polarity inversion line (PIL) \citep{mackay2010}. The observations show that the displacements of the filamentary structures comprising the prominences are mostly aligned with the PIL and hence along the filament axes, without visible changes of the magnetic structure after the onset of the oscillation. In contrast, the magnetic tension drives only transverse motions, which would be perpendicular to the LAL oscillations. \citet{vrsnak2007} assumed a flux-rope geometry, and suggested that these oscillations are driven as a longitudinal-mode standing wave on a tightly coiled spring with fixed ends. This model also predicts motions perpendicular to the local magnetic field, inconsistent with the observed LAL dynamics. Another possibility is that the restoring force is associated with gas pressure differences. However, the temperature differences needed to accelerate the threads to the observed speeds are on the order of several million Kelvins, which are not observed \citep{vrsnak2007}. The damping mechanism is also poorly understood. Several damping mechanisms have been suggested but not rigorously tested, e.g., energy leakage by the emission of sound waves \citep{kleczek1969} or some form of dissipation  \citep{tripathi2009,oliver2009}. 

In this work we propose explanations for the forcing and damping of the LAL oscillations based on our previous investigations of prominence formation and evolution \citep[e.g.,][]{antiochos2000}. We found that all of the resulting condensations oscillate after their formation, and that these motions damp after several oscillations. By studying in detail the motion and damping of many simulated threads in a sheared-arcade filament channel \citep{luna2012}, we could identify and analyze quantitatively the underlying physical processes. Our prominence model thus provides the first self-consistent explanation for the LAL oscillations. 

\section{Numerical model and results}\label{results-sec}

We focus our attention on the $47$ individual oscillating threads modeled in our recent work \citep[for details see][]{luna2012}, which form within dipped flux tubes of a sheared-arcade filament channel \citep{devore2005}; all flux tubes are assumed to remain rigid. The one-dimensional equations governing the plasma dynamics and energetics in these tubes are numerically solved with our ARGOS code \citep{antiochos1999}. In response to steady, localized heating imposed in the tube footpoints, condensations of cool, dense plasma are produced in the corona by the well-investigated thermal nonequilibrium process \citep{antiochos1991,antiochos1999,antiochos2000,karpen2001,karpen2003,karpen2005,karpen2006,karpen2008,xia2011}. 

We found that the combination of heating asymmetry and tube geometry determines the (usually off-center) position where the condensation forms, which is always closer to the footpoint with weaker heating, and the initial velocity of the thread. Threads rapidly settle down after oscillating around the bottom of the dip \citep{antiochos2000,karpen2001,karpen2003}, and accrete mass thereafter at a constant rate due to steady footpoint heating. As discussed in detail below, we find that the oscillatory features of the model prominence threads are strikingly similar to those of the LAL oscillations. The threads move mainly along the filament axis because the local magnetic field is highly sheared (i.e., nearly aligned with the PIL). 

\begin{figure}[!ht]
\centering
\includegraphics[width=9.cm]{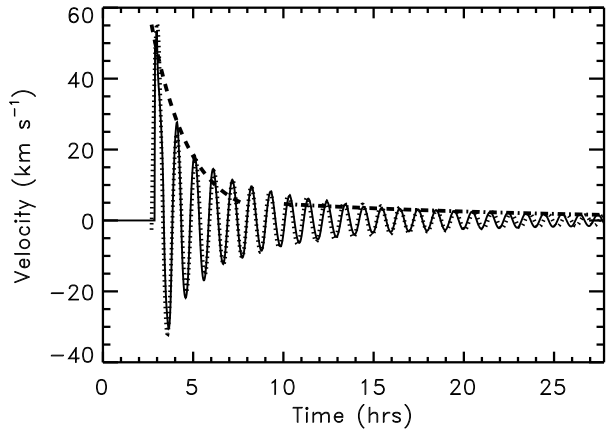}
\caption{Plot of the temporal evolution of the bulk velocity of the center of mass of the condensation example (solid line). The decaying exponential fit for the strong damping (dashed line) and weak damping (dot-dashed line) are also plotted (see text). Additionally, the theoretical velocity fit obtained with our pendulum model described by Eq. \ref{sol-eq} is shown (dotted line).}
\label{osci-tube1}
\end{figure}

A typical example of an oscillating thread is shown in Figure \ref{osci-tube1}. Around $t_\mathrm{c}=2.7~\mathrm{hours}$, the thread forms with an initial velocity amplitude of $49~\mathrm{km~s^{-1}}$ but rapidly reaches a maximum speed of $53~\mathrm{km~s^{-1}}$. The maximum displacement in this case is $40~\mathrm{Mm}$ with respect to the bottom of the dip (the approximate equilibrium position). We note that the oscillation is very regular in time with a constant period of $1.0~\mathrm{hour}$. The oscillation is damped, but the velocity profile cannot be fitted with a sinusoidal function with a single decaying exponential envelope throughout the entire simulation. The initial oscillations damp rapidly at first followed by much weaker attenuation. We have fitted an exponential function for the first three periods of the oscillation envelope (dashed line), yielding strong damping with a decay time of $\tau_\mathrm{s}=2.3~\mathrm{hours}$.  We have also fitted a decaying exponential function for the envelope of the curve after $t=10~\mathrm{hours}$ (dot-dashed line), yielding a weak damping time of $\tau_\mathrm{w}=13.1~\mathrm{hours}$. The existence of these two damping times indicates the existence of two physically distinct phases in the temporal evolution of the thread.

\begin{figure}[!ht]
\centering
\includegraphics[width=9.cm]{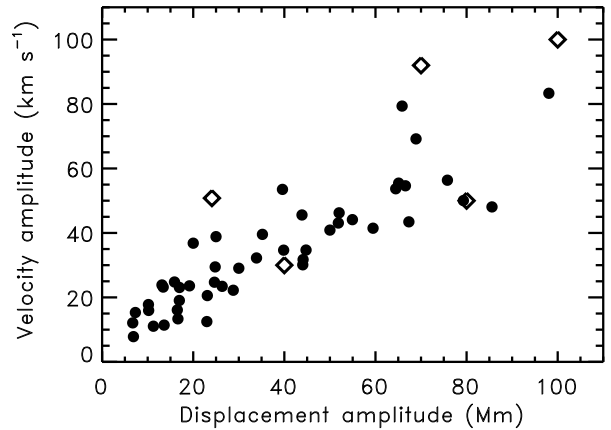}
\caption{Scatter plot of the velocity amplitude as a function of the displacement amplitude for the set of 47 oscillating threads (filled circles), and observational data (diamonds) from \citet{jing2003,jing2006} and \citet{vrsnak2007}.}
\label{amplitudes-fig}
\end{figure}

\begin{figure}[!ht]
\centering
\includegraphics[width=9.cm]{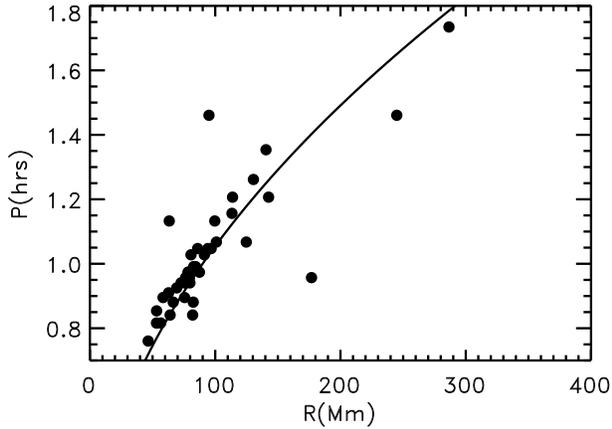}
\caption{Scatter plot of the oscillation period of the threads as a function of the mean radius of curvature of the dipped part of the tubes (filled circles). The solid line shows the theoretical period from the pendulum model described by Eq. \ref{pendulum-frequency-eq}.}
\label{period-fig}
\end{figure}

\begin{figure}[!ht]
\centering
\includegraphics[width=9.cm]{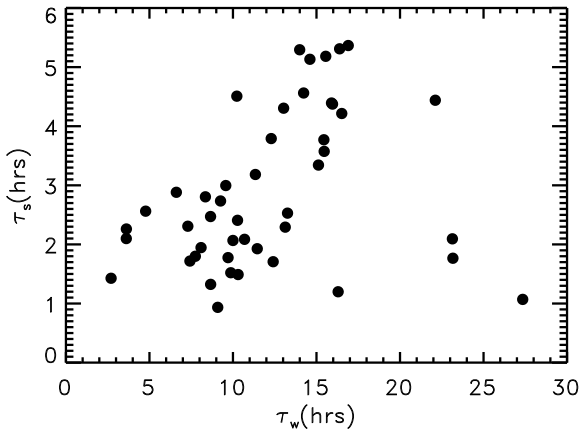}
\caption{Scatter plot of the strong damping time $\tau_\mathrm{s}$ as a function of the weak damping time $\tau_\mathrm{w}$.}
\label{dampings-fig}
\end{figure}

Similarly, we have computed the oscillation parameters for the 47 threads identified in our simulations. Figure \ref{amplitudes-fig} indicates that a higher displacement amplitude is associated with a higher thread velocity. Most of the thread oscillations have velocity amplitudes larger than $20~\mathrm{km}~\mathrm{s}^{-1}$, and hence are classified as ``large amplitude'' oscillations. We have not found any dependence of the periods on the length of the flux tubes or the mass of the condensations. However, we have found a dependence on the average radius of curvature of the dip, $R$, as shown in Figure \ref{period-fig}. In \S \ref{model-sec} we explain the reason for such a dependence. There are few cases that do not fit very well the model curve. These tubes have a irregular shape and the radius of curvature is not uniform in the dip. The strong damping time $\tau_\mathrm{s}$ as a function of the weak damping time $\tau_\mathrm{w}$ is plotted in Figure \ref{dampings-fig}. Clearly $\tau_\mathrm{s} < \tau_\mathrm{w}$ even in the cases where the weak damping time is small. The decay time per period is $\tau_\mathrm{s}/P= 1-6$ for the strong and $\tau_\mathrm{w}/P= 2-27$ for the weak.

The clear agreement between the observed range of amplitudes, periods, and damping times (\S \ref{intro-sec}) and the results of our simulations (see Fig. \ref{amplitudes-fig}) led us to speculate that the reported LAL oscillations could be described with our theoretical model. We propose the following scenario to explain the observed periodic displacements, quasi-aligned with the filament axis, of parts of the filament. The LAL oscillations are triggered by an energetic perturbation at one end of the filament \citep[e.g.,][]{jing2006}; the nature of the trigger is discussed in \S \ref{trigger-sec}. This perturbation pushes the threads to oscillate at the same time, so the LAL oscillations represent the collective motion of bundles of threads that initially move in phase. In our model, the threads oscillate along the dipped part of magnetic flux tubes that are more or less aligned with the axis of the filament. Only a few LAL cycles were observed after the trigger, which we associate with the strong initial damping of the oscillations of our simulated threads. The weak damping has an small contribution to the attenuation in the first cycles. In addition, after several oscillations the excited threads lose their coherence due to slight differences in the oscillation periods, explaining why different parts of the filament oscillate increasingly out of phase as observed by \citet{vrsnak2007}. 

\section{Oscillation physics: restoring force and damping}\label{model-sec}

We noted the existence of thread oscillations in previous numerical studies \citep[e.g.,][]{antiochos2000,karpen2001,karpen2003}, but did not explore the origin of the restoring force and damping. Two possible origins for the restoring force have been proposed: pressure gradients and gravity.  For several reasons, we find that pressure forces cannot account for the observed or simulated behavior. {\it Small-amplitude} prominence oscillations were studied in two-dimensional magnetic configurations with slight flux-tube curvatures by \citet{oliver1992}, \citet{oliver1993}, \citet{joarder1993}, and \citet{oliver1995}, who concluded that the oscillations were pressure driven. However it is unclear whether these linear studies are applicable to large-amplitude, low-frequency oscillations. In the pressure-driven case, rapid changes in thread mass and length would cause frequency shifts \citep[e.g.,][]{morton2009, erdelyi2011} that are not observed in either the solar or the simulated LAL oscillations. Finally, our simulations reveal that the pressure force on the condensation is negligible in comparison with the gravity force. Therefore, we conclude that pressure-driven oscillations are inconsistent with both the LAL observations and with the behavior of our simulated threads. 

We propose instead that the restoring force is mainly the solar gravity, and the damping mechanism is mainly the ongoing accretion of mass by the condensation. In the dipped part of a flux tube, the projected solar gravity is always directed towards the bottom of the dip. Therefore the thread oscillation resembles the motion of pendulum with a length equivalent to the average radius of curvature of the dip, $R$. The temporal evolution of the cool mass \citep[see Fig. 3a of][]{luna2012} consists of an initial transient phase of strong growth starting at time $t_\mathrm{c}$ and a linear regime after $t_{0}$. The transient phase is short $\left(t_{0}-t_\mathrm{c} < P/2\right)$ so the mass of the condensation is well described as
\begin{equation}\label{mass-eq}
m(t)=m_{0}+\alpha~(t-t_0)~,
\end{equation}
where $\alpha$ is the growth rate of the condensation mass, and $m_{0}$ is the mass at the beginning of the linear regime at $t_{0}$. In addition, all parts of the condensation move with similar velocities and the motion is well described by the center-of-mass velocity. Although the threads periodically expand and contract slightly throughout the simulations, these perturbations do not affect the motion of the center of mass. We also assume that the amplitudes of the thread displacements are small with respect to the curvature radii of the dips, consistent with our simulations. The condensation exchanges mass and momentum with the surrounding corona, thus introducing the effects of drag into the force balance. Thus we obtain the equation of motion as
\begin{equation}\label{pendulum-eq}
\frac{d^{2} s}{dt^{2}} + \frac{1}{t-t'} \frac{d s}{dt} + \omega^{2} \,s=0 ~,
\end{equation}
and
\begin{equation}\label{initime-eq}
t'=t_{0}-\frac{m_{0}}{\alpha} =t_{0}-\frac{\psi_{0}}{\omega}~,
\end{equation}
where $s(t)$ is the position of the center of mass of the condensation along the tube, $\omega$ is the angular frequency of the oscillation, and the phase $\psi_{0}=\omega\,m_{0}/\alpha$. Equation (\ref{pendulum-eq}) is similar to the equation of a simple gravity-driven pendulum, with an extra drag term (second term). The coefficient of this drag term is the mass growth rate per unit of mass, $\frac{d m/dt}{m(t)}$. For the initial stages of the evolution ($t \gtrsim t'$) the drag is intense, but for times much larger than $t'$ this term is negligible. 

The angular frequency of the oscillation is
\begin{equation}\label{pendulum-frequency-eq}
\omega= \frac{2\pi}{P}=\sqrt{\frac{g_{0}}{R}}~,
\end{equation}
where $P$ is the period and $g_{0}$ is the gravitational acceleration at the solar surface. Equation (\ref{pendulum-eq}) is the zero-order Bessel equation with solution $J_{0}[\omega (t-t_{0})+\psi_{0}]$. This oscillating function describes very well the first stages of the evolution: the amplitude decreases with time but not exponentially, which explains why we cannot fit a simple decaying exponential to the velocity profile shown in \S \ref{results-sec}. After several periods the Bessel function is basically a sinusoid. However, the simulations also exhibit much slower damping during the later stages of the oscillation (see Fig. \ref{osci-tube1}), so it is necessary to introduce an additional damping term. The thread displacement takes the form
\begin{equation}\label{sol-eq}
s(t)=s_{0} + A\,J_{0}[\omega_{0} (t-t_{0})+\psi_{0}] e^{-(t-t_{0})/\tau_\mathrm{w}}~.
\end{equation}

The velocity derived from this equation is over-plotted in Figure \ref{osci-tube1} (dotted line). We obtain a very good match between the simulated oscillation and the oscillation model, which is also valid for the oscillations of the other $46$ simulated threads. The additional weak damping might be associated with non-adiabatic effects such as conduction and radiation \citep{terradas2001,soler2009}. These losses are very effective in straight flux tubes where the curvature and gravity are not included, and the restoring force is the pressure gradient. In those systems, the non-adiabatic losses weaken the restoring force in each oscillation, damping the movement of the thread. However, in our system the main restoring force is the gravity projected along the tube, and the non-adiabatic losses are not so effective. 

Equation (\ref{sol-eq}) indicates that both mass accretion and non-adiabatic losses contribute to the strong damping at the beginning of the oscillation. However, the former is more important because the latter term has a long damping time. The effective decay time associated with the mass loading is implicit in the Bessel function because its amplitude decreases rapidly in the first few cycles. This damping is related to the phase $\psi_{0}$ of Equation (\ref{sol-eq}). For $\psi_{0}=0$ the damping is maximum and the damping time takes its shortest value, but for larger values of $\psi_{0}$ the damping is weak and the Bessel function is basically a sinusoid. Because this phase is inversely proportional to the mass accretion rate $\alpha$ (see Eq. \ref{initime-eq}), rapid mass loading implies strong damping of the oscillation and vice versa. However, relatively large values of the loading rate imply that $\psi_0 \approx 0$, indicating that the effective damping time is not very sensitive to $\alpha$ variations. The damping depends also on $m_{0}$, which is not straightforward to characterize with a simple analytic approach. For these reasons it is difficult to find a simple relation between the damping time and the parameters of the tube and the heating. This relation will be the subject of a future study.

\section{Prominence seismology}

In our prominence model the magnetic structure exists independent of the condensations. The shear and bending of the coronal magnetic field are determined by the shear at the photosphere and the tension of the overlying field along the PIL \citep{devore2000}. The magnetic-field structure is self supporting, and the interaction with the plasma content is small. However, the magnetic structure must support the denser threads. Thus, the magnetic tension in the dipped part of the tubes must be larger than the weight of the threads or equivalently
\begin{equation}\label{equilibrium-eq}
\frac{B^{2}}{R} - m \,n \,g_{0} \ge 0~,
\end{equation}
where $B$ is the magnetic field strength at the bottom of the dip, $n$ is the particle number density in the thread, the mean particle mass $m=1.27\,m_\mathrm{p}$ \citep{aschwanden2004}, and $m_\mathrm{p}$ is the proton mass. Combining Equations (\ref{pendulum-frequency-eq}) and (\ref{equilibrium-eq}) we obtain
\begin{equation}\label{mag-minimum-eq}
B\ge \sqrt{\frac{g_{0}^{2}\, m \,n}{4\, \pi^{2}}} ~P~,
\end{equation}
which constrains the minimum field strength as a function of the thread oscillation period. Equation (\ref{mag-minimum-eq}) can be written as
\begin{equation}\label{mag-minimum-gauss-eq}
B[\mathrm{\mathrm{G}}] \ge 26\, \left(\frac{n}{10^{11}~\mathrm{cm}^{-3}}\right)^{1/2}\,P[\mathrm{hours}]~.
\end{equation}
For the thread electron number densities of our simulations ($n = 1.6-6.2\times10^{11} ~\mathrm{cm}^{-3}$) and the oscillation periods ($P=0.75-1.75~\mathrm{hours}$), Equation (\ref{mag-minimum-gauss-eq}) yields a minimum field strength of $B \ge 31 - 75~\mathrm{G}$ for the 47 threads with a mean value of $51~\mathrm{G}$,  in agreement with the sparse observations of intermediate-type prominence magnetic fields \citep{mackay2010}. In our prominence model the deep-dip flux tubes have an average area-expansion factor of 3 with respect to their footpoints. Thus we estimate the mean field strength at the flux-tube footpoints as $153~\mathrm{G}$, also consistent with observations \citep[e.g.,][]{aschwanden2004}.

In addition to providing valuable information on the difficult-to-measure coronal magnetic field in filament channels, our model also can determine the radius of curvature of the prominence flux-tube dips from the observed oscillation properties. We have used the \citet{jing2003} and \citet{vrsnak2007} observations to solve Equation (\ref{pendulum-frequency-eq}) for $R$. The observed periods are $80~\mathrm{min}$ and $50~\mathrm{min}$ respectively, yielding corresponding curvature radii of $152~\mathrm{Mm}$ and $62~\mathrm{Mm}$.  This large-scale curvature of the field deduced from the observed oscillations, combined with the large observed thread displacements, argue strongly for the applicability of the sheared-arcade magnetic structure. Assuming that the density of the observed prominence threads is in the range of the values found in our simulations, we estimate the minimum field strength in the prominence as approximately $34-68~\mathrm{G}$ and $23-46~\mathrm{G}$, respectively, for the two cases. 

\section{The oscillation trigger}\label{trigger-sec}

In our simulations, steady footpoint heating produces condensations that oscillate after formation. Each condensation has an initial velocity imparted by the thermal nonequilibrium process and determined by the heating asymmetry and flux tube geometry. However, the observed LAL oscillations clearly are related to an energetic event (subflare, microflare, or flare) near the filament ends. Although a detailed study of the excitation mechanism is beyond the scope of this letter, we propose that the energetic event increases the heating impulsively at one end of the filament, increasing the evaporation of chromospheric plasma into the affected flux tubes. Our preliminary studies have revealed that condensations are remarkably robust; increased heating at one footpoint produces a flow of hot evaporated plasma that pushes a thread to a new location, rather than destroying it {\it in situ}. For example, in the \citet{jing2003} event the subflare pushed an entire section of the filament $70~\mathrm{Mm}$ away from the initial position. We speculate that extra heat is deposited only on those footpoints located closest to the energetic event, so only parts of the filament are excited simultaneously and oscillate collectively in phase. In addition, the increased evaporation produces more mass accretion and increases the oscillation damping. 

\section{Conclusions}\label{summary-sec}

We conclude that the observed LAL motions represent a collective oscillation of a large number of cool, dense threads moving along the magnetic field, triggered by a nearby energetic event (subflare, microflare, or flare). The main restoring force is the projected gravity in the flux tube dips where the threads oscillate, even in threads with large radii of curvature. The period is independent of the tube length and of the constantly growing mass. The oscillations are strongly damped by the mass accretion of the threads. After several oscillations the excited threads lose their coherence due to slight differences in the oscillation periods, causing different parts of the filament to oscillate increasingly out of phase as observed. The mechanism that initiates the collective oscillation is not determined in the present investigation. We speculate that the energetic event impulsively increases the evaporation at the filament flux-tube footpoints closest to the energetic event, increasing the evaporation rate only on one side of the threads. The hot evaporated plasma pushes the existing threads, initiating the oscillations. A detailed study of the initiation mechanism will be the subject of future work. More observations of oscillating prominences are needed to further refine and test our model. 

\acknowledgments

This work has been supported by the NASA Heliophysics SR\&T program. M. L. also acknowledges support from the University of Maryland at College Park and the people of CRESST. Resources supporting this work were provided by the NASA High-End Computing Program through the NASA Center for Climate Simulation at GSFC. We are grateful to our colleagues on international teams on solar prominences hosted by the International Space Science Institute (ISSI) in Bern, Switzerland, especially team leader N. Labrosse, and acknowledge the support of ISSI. We thank H. Gilbert and the referee for constructive comments.

\end{document}